\documentclass[prd,twocolumn,showpacs,showkeys,preprintnumbers,floatfix,superscriptaddress,longbibliography]{revtex4-1}

\usepackage{amsfonts} 
\usepackage{amssymb} 
\usepackage{amsmath} 
\usepackage{graphicx} 
\usepackage{subfigure} 
\usepackage{array} 
\usepackage{dcolumn} 
\usepackage{bm} 
\usepackage{latexsym} 
\usepackage{longtable} 
\usepackage{hyperref} 
\usepackage{bbold}
\usepackage{mathtools}
\usepackage{algorithm}
\usepackage[noend]{algpseudocode}

\usepackage[utf8]{inputenc}
\usepackage[braket]{qcircuit}
\usepackage[dvipsnames,usenames]{xcolor}

\begin{document}
\title{Dynamical Phase Transitions in models of Collective Neutrino Oscillations}

\author{Alessandro Roggero}
\affiliation{InQubator for Quantum Simulation (IQuS), Department of Physics,University of Washington, Seattle, WA 98195, USA}

\preprint{IQuS@UW-21-004}

\date{\today}

\begin{abstract}
Collective neutrino oscillations can potentially play an important role in transporting lepton flavor in astrophysical scenarios where the neutrino density is large, typical examples are the early universe and supernova explosions. It has been argued in the past that simple models of the neutrino Hamiltonian designed to describe forward scattering can support substantial flavor evolution on very short time scales $t\approx\log(N)/(G_F\rho_\nu)$, with $N$ the number of neutrinos, $G_F$ the Fermi constant and $\rho_\nu$ the neutrino density. This finding is in tension with results for similar but exactly solvable models for which $t\approx\sqrt{N}/(G_F\rho_\nu)$ instead. In this work we provide a coherent explanation of this tension in terms of Dynamical Phase Transitions (DPT) and study the possible impact that a DPT could have in more realistic models of neutrino oscillations and their mean-field approximation.
\end{abstract}

\maketitle

When considering astrophysical settings with large neutrino densities, neutrino-neutrino scattering processes can play an important role in shaping the flavor evolution and can lead to collective oscillations in a neutrino cloud~\cite{Pantaleone92,PANTALEONE1992}. This mechanism has been found to play an important role in extreme environments like the early-universe~\cite{Stuart1996,Pastor2002,Abazajian2002} or core-collapse supernovae and binary neutron-star mergers~\cite{Pastor2002B,Balantekin_2005,Fuller2006,Duan2006,Friedland2010,Wu2017,MARTIN2020}. In the latter situations for example, fast neutrino flavor oscillations can lead to important consequences for the revival of the shock wave and nucleosynthesis in the ejected material~\cite{Qian1993,Qian1995,Fogli2007}.

In this work we study simple models of neutrino-neutrino interactions in the forward-scattering limit, when only flavor can be exchanged among neutrinos. For simplicity we also assume that only two flavor of neutrinos mix:  $\nu_e$ corresponding to the electron flavor and $\nu_x$, a combination of $\mu$ and $\tau$ flavors~\footnote{This can be justified if the mixing angle $\theta_{13}=0$ as shown in~\cite{BALANTEKIN1999}.}. In this model, neutrinos are mapped into $SU(2)$ flavor isospins evolving at low densities under the vacuum Hamiltonian~\cite{Pehlivan2011}
\begin{equation}
H_{vac} = \sum_{k=1}^N\frac{\omega_k}{2} \vec{B}_k \cdot \vec{\sigma}_k\;,
\end{equation}
with $\vec{\sigma}_i = (\sigma^x_i,\sigma^y_i,\sigma^z_i)$ the vector of Pauli matrices acting on spin $i$. The one-body coefficients $\omega_k$ are connected to the squared mass gap $\Delta_{m}=m_2^2-m_1^2$ by $\omega_k=\Delta_m/(2E_k)$, with $E_k$ the neutrino energy. The neutrino mass hierarchy is reflected in the sign of the gap: for normal hierarchy we consider $\Delta_{m}>0$, while for inverted hierarchy we take $\Delta_m<0$~\cite{Duan2010,Pehlivan2011}.
The orientation of the "magnetic field" vector $\vec{B}_k=(\sin(2\theta),0,-\cos(2\theta))$ is related to the mixing angle $\theta$. Importantly, the collective oscillations discussed in this work are not related to the presence of off-diagonal components in the $H_{vac}$ Hamiltonian and in order to avoid confusion we will use a global $SU(2)$ rotation to move to the mass basis $\ket{\downarrow}=\ket{\nu_x}$ and $\ket{\uparrow}=\ket{\nu_e}$ with a diagonal vacuum Hamiltonian.

With the addition of the forward-scattering weak interaction among neutrinos, the full Hamiltonian reads~\cite{Pehlivan2011}
\begin{equation}
\label{eq:fs_hamilt}
H_{FS} =- \sum_{k=1}^N \frac{\omega_k}{2} \sigma^z_k + \frac{\mu}{2N}\sum_{i<j}^N \mathcal{J}_{ij} \vec{\sigma}_i\cdot\vec{\sigma}_j\;,
\end{equation}
where the interaction strength is given by $\mu=\sqrt{2}G_F\rho_\nu$, with $G_F$ the Fermi constant and $\rho_\nu$ the neutrino number density.
The geometry of the problem is encoded in the coefficients of the two-body coupling matrix $\mathcal{J}_{ij}$ as
\begin{equation}
\label{eq:jcoupling}
\mathcal{J}_{ij} = \left(1-\frac{\vec{p}_i\cdot\vec{p}_j}{|\vec{p}_i||\vec{p}_j|}\right) = \left(1-\cos(\theta_{ij})\right)\;,
\end{equation}
with $\vec{p}_k$ the momentum associated with the $k$-th neutrino.

In the low density limit $\mu\ll\omega_k$, the neutrinos oscillate independently with their own frequency $\omega_k$. The presence of the forward-scattering interaction can allow collective effects to develop when $\mu\gtrsim\omega_k$ giving rise to interesting phenomena like synchronization~\cite{Pastor2002,Fuller2006,Raffelt2010,AKHMEDOV2016}, bipolar oscillations~\cite{Kostelecky1995,Duan2006b,Duan2007b} and spectral splits/swaps~\cite{Duan2006c,Duan2007,Raffelt2007,Dasgupta2009,Martin2020b}. Due to the computational complexity of solving directly for the dynamics generated by the Hamiltonian $H_{FS}$ for large systems, much of the current understanding of collective oscillation phenomenology is derived within mean-field approaches (see~\cite{Duan2010} for a review) which, owing to the infinite range of the interaction in Eq.~\eqref{eq:fs_hamilt}, are expected to become increasingly correct as we approach the thermodynamic limit $N\gg1$ (this is true in general at least for the ground-state energy, see eg.~\cite{Brandao2016}). 

In this work we are interested in understanding the out-of-equilibrium dynamics of the spin model for large but finite systems in order to understand the rate of convergence to the mean field result. Early work by Friedland and Lunardini~\cite{Friedland2003} studied the Hamiltonian in Eq.~\eqref{eq:fs_hamilt} in the limit where the vacuum term is negligible (high density) and assuming the geometry is isotropic.  In this limit, the Hamiltonian is proportional to the total angular momentum operator and therefore easily diagonalizable. The exact solution shows that substantial flavor evolution occurs only for the time scales $\tau_L\approx\mu^{-1}\sqrt{N}$ associated with incoherent scattering. The result is fully consistent with Ref.~\cite{Friedland2003b} which argued, using a short-time approximation, that no entanglement is generated in the many-body evolution of the system and that the mean-field picture of incoherent scattering is correct.

The original study in Ref.~\cite{Friedland2003} was motivated by earlier work by Bell, Rawlinson and Sawyer~\cite{Bell2003} which presented numerical evidence from a similar model, where however $SU(2)$ invariance was explicitly broken, supporting a very different result: neutrino flavor evolution occurring on much shorter time scales $\tau_S=\mu^{-1}$, independently of system size. Despite the infinite range of the pair interaction in $H_{FS}$, one can expect the time for information to propagate throughout the whole system to be lower bounded by the information signaling time scaling as $\tau_{si}\propto\log(N)$ instead (see eg.~\cite{Guo2020}). Later work by Sawyer~\cite{sawyer2004classical} provided additional numerical evidence, with larger system sizes, suggesting indeed the presence of collective flavor oscillations on a fast time scale $\tau_F\approx\mu^{-1}\log(N)$. 

In the present work, we propose an explanation for the emergence of these different time scales, in apparently very similar models for the neutrino forward scattering problem, as a consequence of the presence of a Dynamic Phase Transition (DPT)~\cite{Heyl2013,Heyl2018} in the spin system. The models considered in~\cite{Bell2003,sawyer2004classical}, and described in more detail in Sec.~\ref{sec:high_dens} below, give rise to fast oscillations with times scaling as $\tau_F$ by introducing however an unphysical perturbation that breaks the $SU(2)$ invariance of the neutrino Hamiltonian in Eq.~\eqref{eq:fs_hamilt}. As shown recently in a companion paper~\cite{Roggero2021A}, the presence of of the vacuum Hamiltonian $H_{vac}$ can also produce fast oscillations with times scaling as $\tau_F$. In Sec.~\ref{sec:intdens} we provide additional details about these results and establish a stronger connection with the underlying DPT.
Finally, we provide a summary and conclude in Sec.~\ref{sec:conc}.

\section{High density limit}
\label{sec:high_dens}

It is reasonable to expect that collective effect would be enhanced in the high density limit where $\mu\gg1$ and the neutrino-neutrino coupling is strong. In the next two sections we will study neutrino systems in the limit where $\mu\gg|\omega_k|$ and neglect the vacuum one-body part from the full Hamiltonian. This contribution will be reintroduced and shown to play an important role in Sec.~\ref{sec:intdens} below.

\subsection{Single angle approximation}
\label{sec:sa}

We start our discussion with the model obtained using a very common simplification: the single angle approximation. This amounts to neglect the spatial information encoded in the coupling matrix $\mathcal{J}_{ij}$ from Eq.~\eqref{eq:jcoupling} and replace it with it's average value. Here and in the following we will take, without loss of generality, the coupling to be $\mathcal{J}_{ij}=1$. The final Hamiltonian, after neglecting the one-body vacuum term, can then be written as
\begin{equation}
\label{eq:sa_ham}
H_{sa} = \frac{\mu}{2N} \sum_{i<j} \vec{\sigma}_i \cdot\vec{\sigma}_j = \frac{\mu}{N}J^2 - \frac{3}{4}\mu\;,
\end{equation}
where we introduced the total flavor spin $\vec{J}=\frac{1}{2}\sum_i\vec{\sigma}_i$. This model is similar to the Lipkin-Meshov-Glick (LMG) model~\cite{Lipkin1965} which, together with it's variants, has been explored extensively in the past~\cite{Vidal2004,Vidal2004b,Vidal2004c,Latorre2005,Ribeiro2008}. The Hamiltonian $H_{sa}$ is diagonal in the angular momentum basis $\ket{j,m}$ with $j\in{0,\dots,N/2}$ and eigenvalues given by
\begin{equation}
E_{sa}(j,m) = \frac{\mu}{N} j\left(j+1\right) - \frac{3}{4}\mu\;.
\end{equation}
The ground-state is the singlet $\ket{0,0}$ and the gap to excited states with total spin less than $N^{\eta/2}$ vanishes in the thermodynamic limit for any $\eta<1$.
Owing to the high degree of symmetry of this model, analytical solutions can be found for the evolution of any observable quantity as a function of time. In particular, a useful observable considered also in Refs.~\cite{Friedland2003,Bell2003,Roggero2021A} is the flavor persistence $p(t)$, defined as the probability of measuring one of the neutrinos in the same flavor state it had at the beginning of time evolution. Throughout this work we will consider an initial product state defined as
\begin{equation}
\label{eq:init_state}
\ket{\Psi_0} = \left(\bigotimes_{n=1}^{N/2} \ket{\downarrow} \right)\otimes\left(\bigotimes_{m=1}^{N/2} \ket{\uparrow} \right)\;.
\end{equation}
In this case the flavor persistence can be expressed explicitly as the following expectation value
\begin{equation}
\label{eq:flav_pers}
p(t) = \frac{1}{2}\langle\Psi(t)\lvert(1-\sigma^z_1)\rvert\Psi(t)\rangle\;.
\end{equation}
where $\ket{\Psi(t)}=\exp(-itH)\ket{\Psi_0}$ is the time evolved state and, without loss of generality, we have considered the first neutrino which started in the heavy flavor state $\ket{\downarrow}$ at time $t=0$.
Here and in the following, we will denote the two sets of spins initialized with opposite polarizations in $\ket{\Psi_0}$ as $A$ and $B$, with corresponding total spin operators $\vec{J}_A=\left(X_A,Y_A,Z_A\right)$ and $\vec{J}_B$ respectively.

In order to expose the role of Dynamical Phase Transitions in the collective oscillation phenomenon, we want to describe the full time evolution of the initial state $\ket{\Psi_0}$ under the Hamiltonian in Eq.~\eqref{eq:sa_ham} as a quantum quench~\cite{Polkovnikov2011}. In this setup one starts with an initial Hamiltonian $H^0_{sa}$, of which $\ket{\Psi_0}$ is a ground state of, and suddenly changes to the final Hamiltonian $H_{sa}$ given above. With our choice of initial state $\ket{\Psi_0}$, the initial Hamiltonian we consider in this case can be chosen as
\begin{equation}
\label{eq:init_ham_sa}
H^0_{sa} = \frac{\nu}{4N} \sum_{i\in\mathcal{A}}\sum_{i\in\mathcal{B}}\sigma^z_i \sigma^z_j = \frac{\nu}{N} Z_AZ_B\;,
\end{equation}
where we have indicated with $\mathcal{A}$ and $\mathcal{B}$ the set of indices for the spins of the $A$ and $B$ group. The Hamiltonian $H^0_{sa}$ has two degenerate ground-states corresponding to $\ket{\Psi_0}$ and to it's spin-reversed partner obtained by applying the Pauli $X$ operator to each spin: $\ket{\Psi_1}=\bigotimes_{i}\sigma^x_i\ket{\Psi_0}$.

The full Hamiltonian used for our quantum quench can then be express compactly as follows
\begin{equation}
\label{eq:full_ham}
H(t) = \frac{\mu(t)}{N}J^2  + \frac{\nu(t)}{N} Z_AZ_B\;,
\end{equation}
with $(\mu(0),\nu(0))=(0,1)$ and $(\mu(t),\nu(t))=(1,0) \;\forall t>0$. The system described by the full Hamiltonian $H(t)$ undergoes a quantum phase transition between a gapped phase for $\nu(t)\gg\mu(t)$ to a gapless phase for $\nu(t)\ll\mu(t)$. Contrary to the gapless Hamiltonian $H_{sa}$, the full Hamiltonian $H(t)$ in Eq.~\eqref{eq:full_ham} is not diagonal in the coupled angular momentum basis $\ket{J,M}$. Using a mean-field calculation, which is exact in the thermodynamic limit, we find for $\mu>0$ a critical point at $\nu=0$ in the thermodynamic limit (see Appendix~\ref{app:phase_sa} for more details). The quench dynamics under consideration here will therefore terminate at the quantum critical point.

In order to define and characterize in general a Dynamical Phase Transition (see~\cite{Heyl2018} for a review) one usually starts by introducing the Loschmidt echo as
\begin{equation}
\label{eq:loch}
\mathcal{L}(t) = \left|\langle \Phi\lvert \exp\left(-i t H_f\right)\rvert\Phi\rangle \right|^2\;,
\end{equation}
with $\ket{\Phi}$ the initial (pure) state at $t=0$ and $H_f$ the final Hamiltonian of the quench. The quantity $\mathcal{L}(t)$ is a fidelity measure~\cite{Gorin2006} that quantifies the probability for the system to return to it's initial state. A DPT is then characterized by non-analiticities in the rate function
\begin{equation}
\label{eq:loch_rate}
\lambda(t) = -\frac{1}{N}\log\left[\mathcal{L}(t)\right]\;,
\end{equation}
where $N$ is the total number of particles in the system and $\lambda(t)$ an intensive "free energy"~\cite{Heyl2013,Gambassi2012}. The rate $\lambda(t)$ plays here the role of a non-equilibrium equivalent of the thermodynamic free-energy. Notably, other definitions of DPT are possible, for instance using time averaged order parameters~\cite{Sciolla2011,Sciolla2013,Zunkovic2018} and there are known cases where the two definitions of criticality are incompatible~\cite{Zunkovic2016}. In the rest of this work we consider only DPT characterized using the Loschmidt echo and leave for future work a more detailed connection to dynamical order parameters.

Due to the degeneracy in the ground-space of the initial Hamiltonian $H^0_{sa}$, the Loschmidt echo in Eq.~\eqref{eq:loch} needs to be generalized. As shown in Refs.~\cite{Heyl2014,Zunkovic2018} a consistent generalization can be found by considering the total probability $P(t)$ of returning to the ground-space
\begin{equation}
\label{eq:fullp}
P(t) = \mathcal{L}_0(t) + \mathcal{L}_1(t)\;,
\end{equation}
where we introduced the two Loschmidt echoes
\begin{equation}
\label{eq:two_echoes}
\mathcal{L}_k(t) = \left|\langle \Psi_k\lvert \exp\left(-it H_{sa}\right)\rvert\Psi_0\rangle \right|^2\;,
\end{equation}
associated with both ground-states. In the thermodynamic limit $N\gg1$ only one of the two contribution will dominate resulting in the asymptotic scaling~\cite{Heyl2014}
\begin{equation}
P(t) \to e^{-N\lambda_m(t)}\quad\lambda_m(t)=\min\left[\lambda_0(t),\lambda_1(t)\right]\;,
\end{equation}
up to exponentially small corrections. The rate functions $\lambda_0(t)$ and $\lambda_1(t)$ correspond to the definition in Eq.~\eqref{eq:loch_rate} but applied to $\mathcal{L}_0(t)$ and $\mathcal{L}_1(t)$ separately. A DPT can then occur whenever $\mathcal{L}_0(t)$ and $\mathcal{L}_1(t)$ intersect at some finite value $t^*$ for the evolution time~\cite{Heyl2014,Zunkovic2018}.
According to the phase diagram described above, our initial state is quenched up to the critical point and this could lead to a finite value of the crossing time $t^*$ for any finite $N$. 

\begin{figure}
 \centering
 \includegraphics[width=0.49\textwidth]{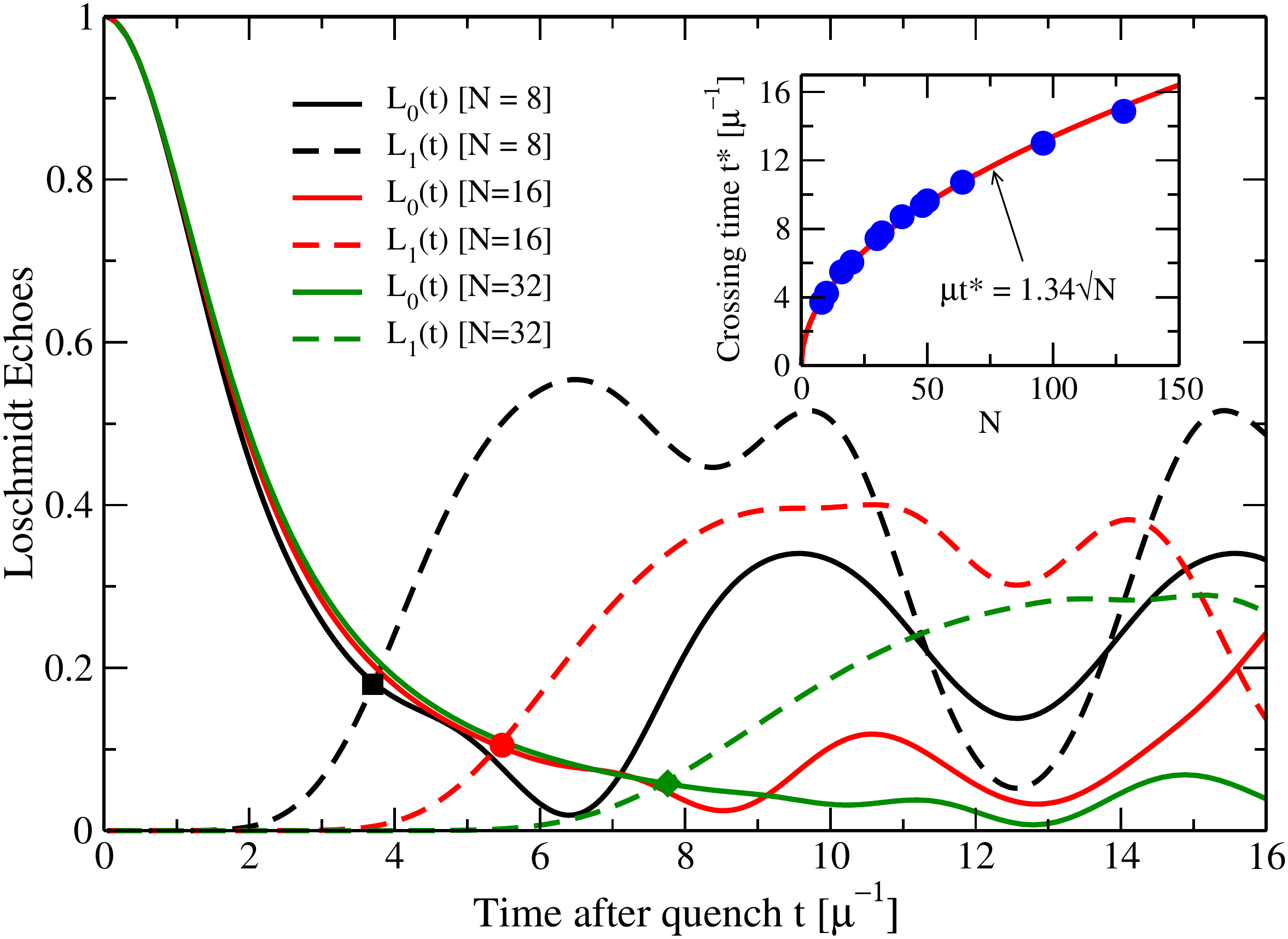}
 \caption{(Color online) Time evolution of the Loschmidt echoes $\mathcal{L}_0(t)$ and $\mathcal{L}_1(t)$ for different systems sizes $N$. The inset shows the crossing time as a function of system size.}
\label{fig:sa_echoes}
\end{figure}

In order to test this scenario, we performed numerical simulations using the Time Evolving Block Decimation (TEBD) algorithm with Matrix Product States (MPS)~\cite{Vidal2003_mps} implemented using the iTensor library~\cite{itensor}. The appealing property of this class of algorithms is that their computational cost scales with the amount of entanglement generated by the real-time dynamics and can then be used efficiently when quantum correlations are sufficiently weak. The implementation of the time evolution operator $U(t)=\exp(-itH)$ follows the swap network scheme employed also in past quantum simulations~\cite{Hall2021}. 
Additional details on this computational scheme can be found in the companion paper Ref.~\cite{Roggero2021A}.

The results in the main panel of Fig.~\ref{fig:sa_echoes} show the two Loschmidt echoes $\mathcal{L}_0(t)$ and $\mathcal{L}_1(t)$ for system of different size. In marked difference with the nearest neighbour case studied in Ref.~\cite{Heyl2014}, the crossing time $t^*$ shows a rapid evolution with system size on time scales proportional to $\tau_L=\mu^{-1}\sqrt{N}$. From the results of our simulations we extract a value of $t^*/\tau_L= 1.34(2)$ for the crossing time. The divergence of $\tau_L$ with the system size $N$ indicates that this is not technically a DPT, in the sense that the crossing of Loschmidt echoes is a finite-size effect that will vanish in the thermodynamic limit.

\begin{figure}[t]
 \centering
 \includegraphics[width=0.49\textwidth]{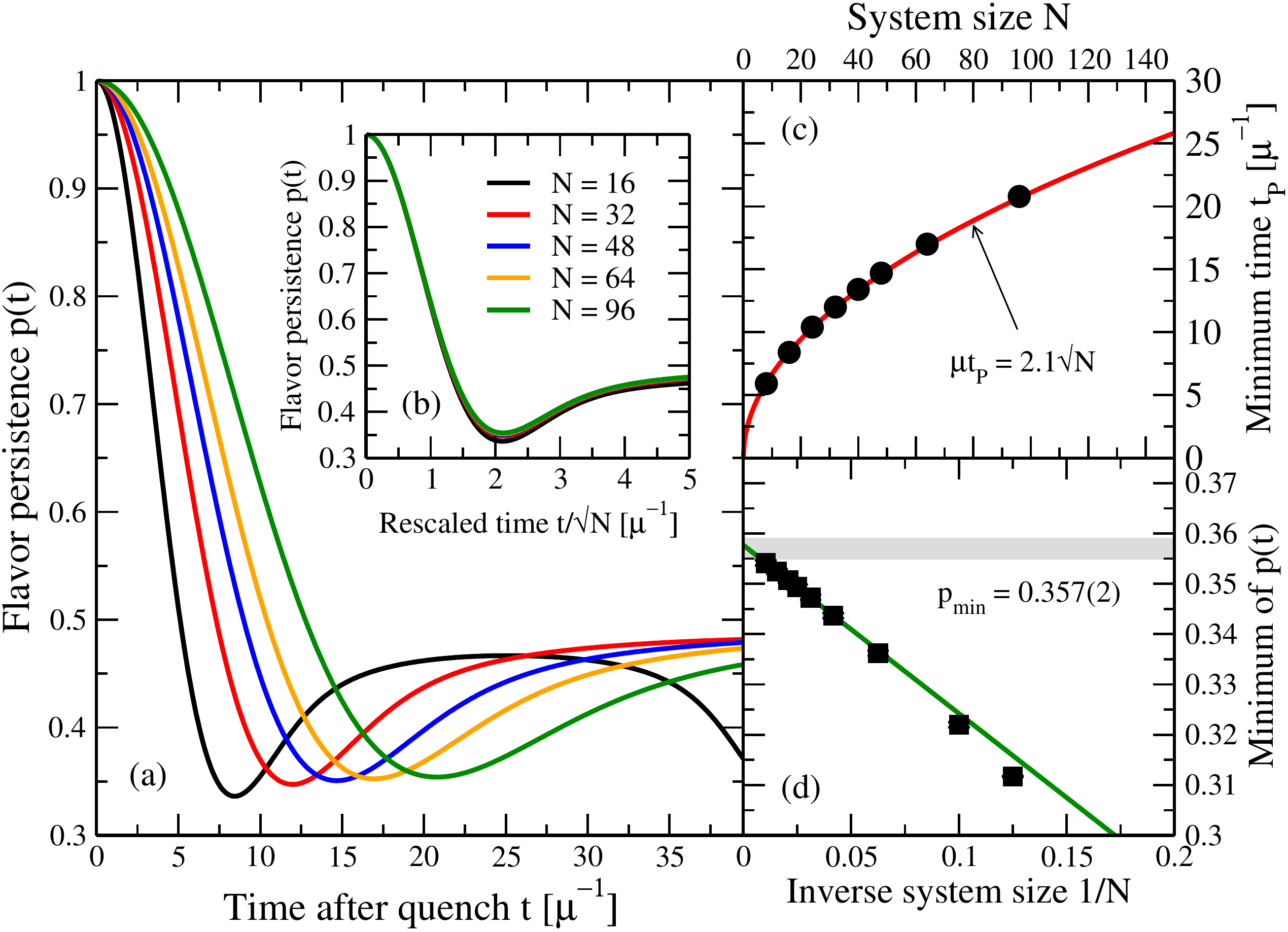}
 \caption{(Color online) The main panel (a) shows the time evolution of the flavor persistence $p(t)$ for different system sizes $N=[16,32,48,64,96]$. The inset (b) shows the persistence $p(t)$ plotted versus the rescaled time $t'=t/\sqrt{N}$. The data in panel(c) shows the evolution with system size of the time $t_P$ to reach the minimum of $p(t)$ while in panel (d) we report the value of the persistence at the minimum as a function of $1/N$. The continuous curves in panels (c) and (d) correspond to the fit described in the text.}
\label{fig:sa_pers}
\end{figure}

The results of our simulation for the the flavor persistence $p(t)$, defined explicitly in Eq.~\eqref{eq:flav_pers}, are shown in Fig.~\ref{fig:sa_pers}. We recover the result reported in Ref.~\cite{Friedland2003}: the minimum of the persistence is achieved at times $t_P\propto\tau_L$. This is clearly indicated by the inset (b) of Fig.~\ref{fig:sa_pers} which shows the persistence as a function of the rescaled time $t'=t/\sqrt{N}$. The dependence on system size is minimal.

The right hand panels show more in detail the system size dependence of $t_P$, in panel (c), and of the value $p_{min}(N)$ of the persistence at it's minimum, in panel (d). The latter is plotted as a function of $1/N$ to emphasize the power law scaling of $p_{min}(N)=p_{min}-c/N$ (the solid green curve in panel (d)). The result of these fit for the minimum time is $t_P/\tau_L = 2.10(5)$ while $p_{min}=0.357(2)$ in the infinite system size limit. The first two data points in panel (d) of Fig.~\ref{fig:sa_pers} correspond to $N=8$ and $N=10$ and we see that one needs to reach $N=16$ before deviations from the $1/N$ behavior are apparent.

All of these time scales quickly diverge for large system sizes $N\gg1$ and the mean-field solution, which predicts for this models no time evolution at all, becomes eventually exact in the thermodynamic limit.

In order to quantify quantum correlations in the evolved state, we compute the half-chain entanglement entropy (see eg.~\cite{Eisert2010}) defined as
\begin{equation}
\label{eq:entropy}
S_{N/2}(t) = -Tr\left[\rho_\mathcal{B}(t)\log_2\left(\rho_\mathcal{B}(t)\right)\right]\;,
\end{equation}
with $\rho_\mathcal{B}=Tr_{\mathcal{A}}\left[\rho(t)\right]$ the reduced density matrix obtained by tracing the full density matrix of the neutrino system at time $t$, denoted as $\rho(t)$, over the first $N/2$ spins belonging to the $\mathcal{A}$ group defined above. 

\begin{figure}[tb]
 \centering
 \includegraphics[width=0.49\textwidth]{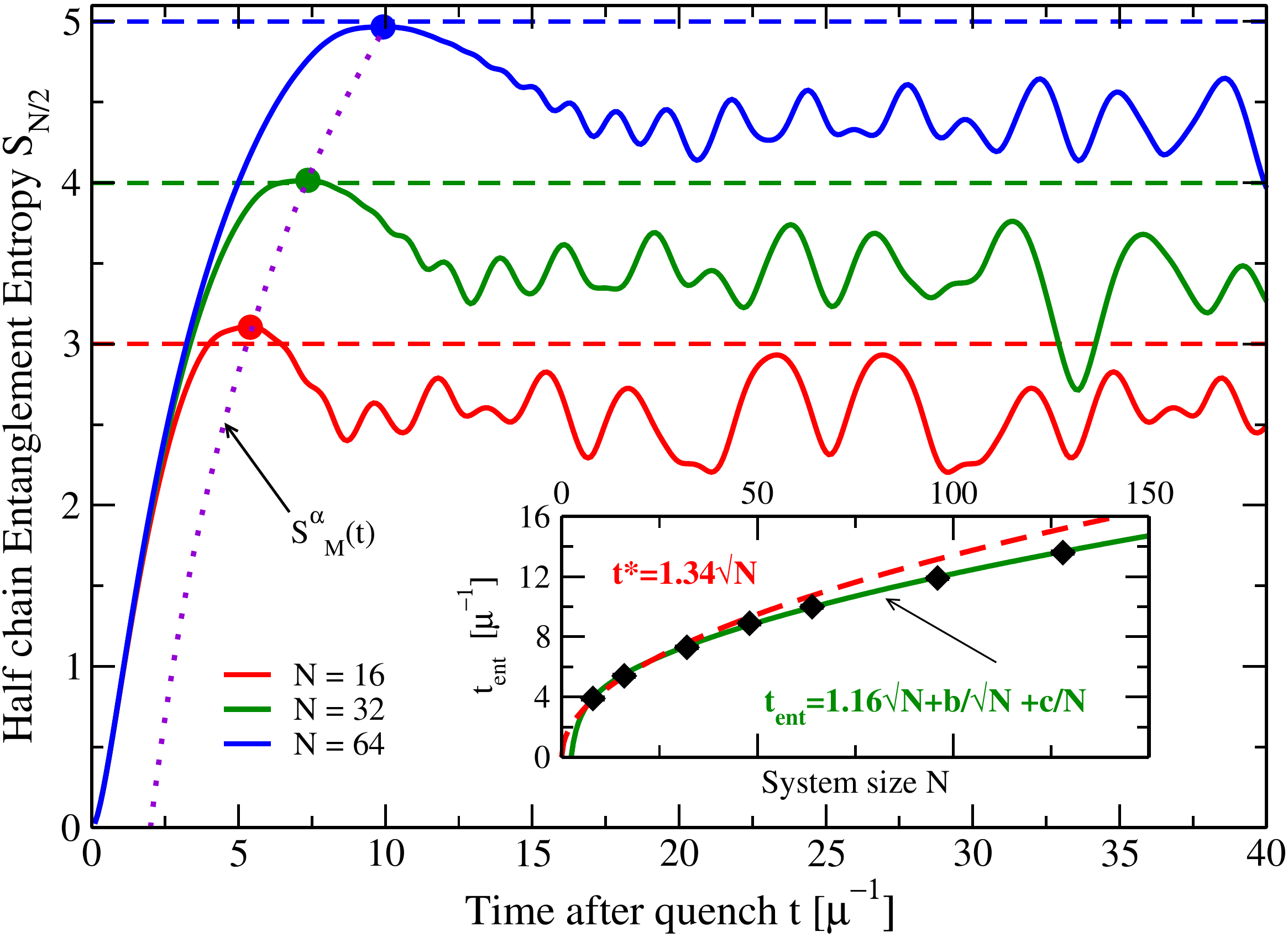}
 \caption{(Color online) Time evolution of the half-chain entropy. Horizontal dashed lines correspond to the value $\log_2(N/2)$ and the (purple) dotted line is the fit from Eq.~\eqref{eq:smaxoft}. The inset shows the evolution of the time to reach maximum entropy with system size: the green solid curve is the best fit discussed in the text while the red dashed line corresponds to the crossing time $t^*$ from Fig.~\ref{fig:sa_echoes}.}
\label{fig:sa_entropy}
\end{figure}

We see from the results in Fig.~\ref{fig:sa_entropy} that, after an initial growth, the entropy $S_{N/2}(t)$ reaches a peak and then plateaus at a value $S_{max}\approx\log_2(N/2)$ with oscillations around the average. The maximum value $S_{max}$ for the entanglement entropy is reminiscent to the one in ground states of one dimensional spin systems at a quantum critical point~\cite{Vidal2003,Refael2004} and reflects the absence of a gap in the Hamiltonian $H_{sa}$ in Eq.~\eqref{eq:sa_ham}. The qualitative behavior of $S_{N/2}(t)$ is remarkably close to the one observed with a similar model (but different initial conditions) in Ref.~\cite{Pappalardi2018} where the entanglement entropy was observed to peak and then plateau when the system was quenched at the critical point of a DPT. The observed time scale to reach the peak, also connected to the Eherenfest time $t_{Ehr}$~\cite{Pappalardi2018}, was found there to scale as $t_{Ehr}\approx\log(N)$ similarly to the fast scale $\tau_F$ while away from the quantum critical point $t_{Ehr}\approx\sqrt{N}$ like $\tau_L$.

From our simulation we find that in our case, despite being at the critical point, the entropy grows more slowly and reaches the peak on the slow time scale $t_{ent}\approx \sqrt{N}$. In order to account for finite size effects, we perform a fit to the data shown in the inset of Fig.~\ref{fig:sa_entropy} using
\begin{equation}
t_{ent}(N) = a \sqrt{N} + \frac{b}{\sqrt{N}} + \frac{c}{N}
\end{equation}
The optimal parameter for the leading order term is found to be $a\tau_L = 1.16(4)$ while the finite size corrections $b$ and $c$ are $\mathcal{O}(10)$. This time scale is very similar, and always strictly smaller, to the crossing time $t^*$ when the DQPT occurs (see red dashed line in inset of Fig.~\ref{fig:sa_entropy}).

A separate test of whether the entanglement time $t_{ent}$ scales algebraically (case $\alpha$) or logarithmically (case $\beta$) in system size can be obtained by estimating the time to reach $S_{N/2}=S_{max}$ using two limiting cases
\begin{equation}
\label{eq:smaxoft}
S_M(t) = \log_2\left(\frac{N(t)}{2}\right) =\bigg{\{} \begin{matrix}A\log_2(t/B)&\text{case }\alpha\\
Ct+D&\text{case }\beta\\
\end{matrix}\;.
\end{equation}

For the single angle setup considered in this section we found a good fit to data only for the model from "case $\alpha$" (shown as purple dotted line in Fig.~\ref{fig:sa_entropy}) with optimal parameters $A=2.14(4)$ and $B=2$ respectively.

We note that this slow increase of the entanglement entropy with system size and with time is at the hearth of the classical simulatability of the neutrino model in the single-angle approximation with Matrix Product States: the maximum bond dimension needed only scales linearly with $N$ to obtain converged results. The TEBD scheme employed here, and in the accompanying paper~\cite{Roggero2021A}, is however not optimal for long range interactions and further progress could be made using more sophisticated simulation techniques like the Time Dependent Variational Principle~\cite{Haegeman2011} as well as different tensor network~\cite{Vidal2008,Evenbly2011} or neural network states~\cite{Deng2017}.

\subsection{Fast oscillations with $SU(2)$ breaking}
\label{sec:su2b}

The first calculations showing a many-body "coherent speedup" of flavor oscillations at the shorter time-scale $\tau_S=\mu^{-1}$ were obtained in Refs.~\cite{Bell2003,sawyer2004classical} using a neutrino Hamiltonian that explicitly breaks the global $SU(2)$ flavor invariance of the Hamiltonian $H_{FS}$ in Eq.~\eqref{eq:fs_hamilt}. The symmetry-breaking term used in both cases is
\begin{equation}
H_{SB} = \left(\Delta-1\right)\frac{\mu}{2N}\sum_{i<j}^N \mathcal{J}_{ij}\sigma^z_i\sigma^z_j\;.
\end{equation}
The control parameter here is $\Delta$ and for $\Delta=1$ the original $SU(2)$ invariant interaction is recovered. As we will see below the geometry of the problem encoded in the angular factors $\mathcal{J}_{ij}$ will play now an important role. 

In this section we will consider a very simple situation: two neutrino beams, one with $N/2$ neutrinos starting in the $\ket{\downarrow}$ state and one with $N/2$ neutrinos starting in $\ket{\uparrow}$. These correspond to the sets $\mathcal{A}$ and $\mathcal{B}$ defined above. Neutrinos belonging to the same group interact with the same strength $\mathcal{J}_{AA}=\mathcal{J}_{BB}$ while neutrinos belonging to different beams interact with a coupling $\mathcal{J}_{AB}$. Using the total flavor spin operators $\vec{J}_A$ and $\vec{J}_B$ introduced above, we can write the full Hamiltonian used in this quench as
\begin{equation}
\begin{split}
\label{eq:ham_tb}
H_{tb} &= \frac{\mu \mathcal{J}_{AA}}{N}\left[ J_A^2 + J_B^2 + \left(\Delta-1\right) \left(Z_A^2+Z_B^2\right)\right]\\
&+\frac{2\mu \mathcal{J}_{AB}}{N} \left[\vec{J}_A\cdot\vec{J}_B +\left(\Delta-1\right)Z_AZ_B \right]\;,
\end{split}
\end{equation}
plus an inconsequential constant factor that we ignore. The limit in which the beams are very collimated 
corresponds to the choice $\mathcal{J}_{AA}=0$, and the weak interactions are relevant only across beams. For the rest of this section we will measure energies in units of $(\mu\mathcal{J}_{AB})$ and use directly the dimensionless parameter $\Gamma=\mathcal{J}_{AA}/\mathcal{J}_{AB}$.

The quench dynamics we will consider in this section starts in the limit $\Delta\to\infty$ with $\Gamma\to0$ which corresponds to the starting Hamiltonian $H^0_{sa}$ considered above, with $\ket{\Psi_0}$ and $\ket{\Psi_1}$ as it's two degenerate ground-states.

\begin{figure}
 \centering
 \includegraphics[width=0.49\textwidth]{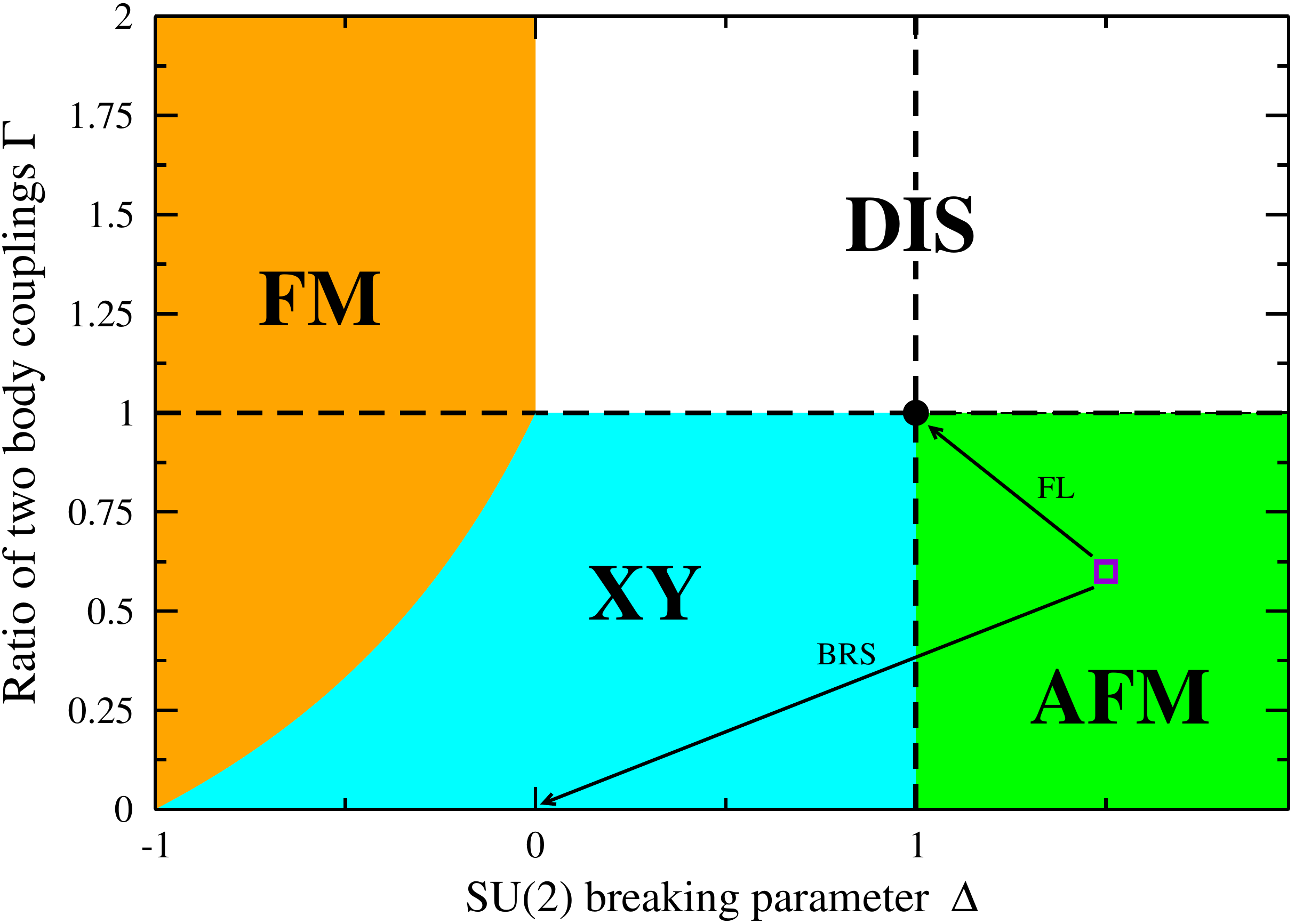}
 \caption{(Color online) Equilibrium phase diagram for the two beam model (see App.~\ref{app:tb_phase_diag} for a derivation) together with the two quantum quenches considered in the main text: FL denotes the single angle model from Ref.~\cite{Friedland2003}, BRS indicates the $SU(2)$ broken model from Ref.~\cite{Bell2003}. In both cases the system starts with $\ket{\Psi_{0}}$ in the AFM phase (purple square). The dashed lines indicate the set of points in parameter space where the dynamics is equivalent to that of the single angle $SU(2)$ invariant point (denoted by a solid circle).}
\label{fig:2b_quench}
\end{figure}

The equilibrium phase diagram of the two beam Hamiltonian $H_{tb}$ is now much richer than with the single angle approximation (see Fig.~\ref{fig:2b_quench}). Using a mean-field approach (see App.~\ref{app:tb_phase_diag} for details) we can identify 4 distinct phases depending on the value of the $SU(2)$ breaking parameter $\Delta$ and on the ratio $\Gamma$ of the two body couplings which specifies the relative orientations of the beams.

For collimated beams with $\Gamma<1$ we find two gapped phases, one with anti-ferromagnetic order in the z direction at large positive values of $\Delta$ (denoted by $AFM$) and one with ferromagnetic order along the $z$ direction for sufficiently negative values of $\Delta$ (denoted by $FM$). These two phase are separated by a gapless phase, indicated by $XY$ in Fig.~\ref{fig:2b_quench}, where anti-ferromagnetic order is preserved in the $xy-$plane but is lost in the $z$ direction. The only ordered phase present for $\Gamma\geq1$ is the $FM$ phase for $\Delta<0$ while a disordered gapless phase emerges for positive values of $\Delta$ (denoted by $DIS$ in Fig.~\ref{fig:2b_quench}).

The results within the single angle approximation described in the previous section (and in Ref.~\cite{Friedland2003}) correspond to the trajectory indicated by the $FL$ arrow in Fig.~\ref{fig:2b_quench} and ending at the full dot (which indicates the single angle point). The dashed lines emanating from that point indicate parameter values for which, due to conservation laws, the dynamics is indistinguishable from the one obtained with the $FL$ quench. Note that this holds also for quenches that are apparently crossing a phase boundary. This is in agreement with previous studies showing that a DPT can fail to appear even in quenches that crossed a phase boundary (see eg.~\cite{Vajna2014,Sharma2015,Zunkovic2018}).

\begin{figure}[bh]
 \centering
 \includegraphics[width=0.49\textwidth]{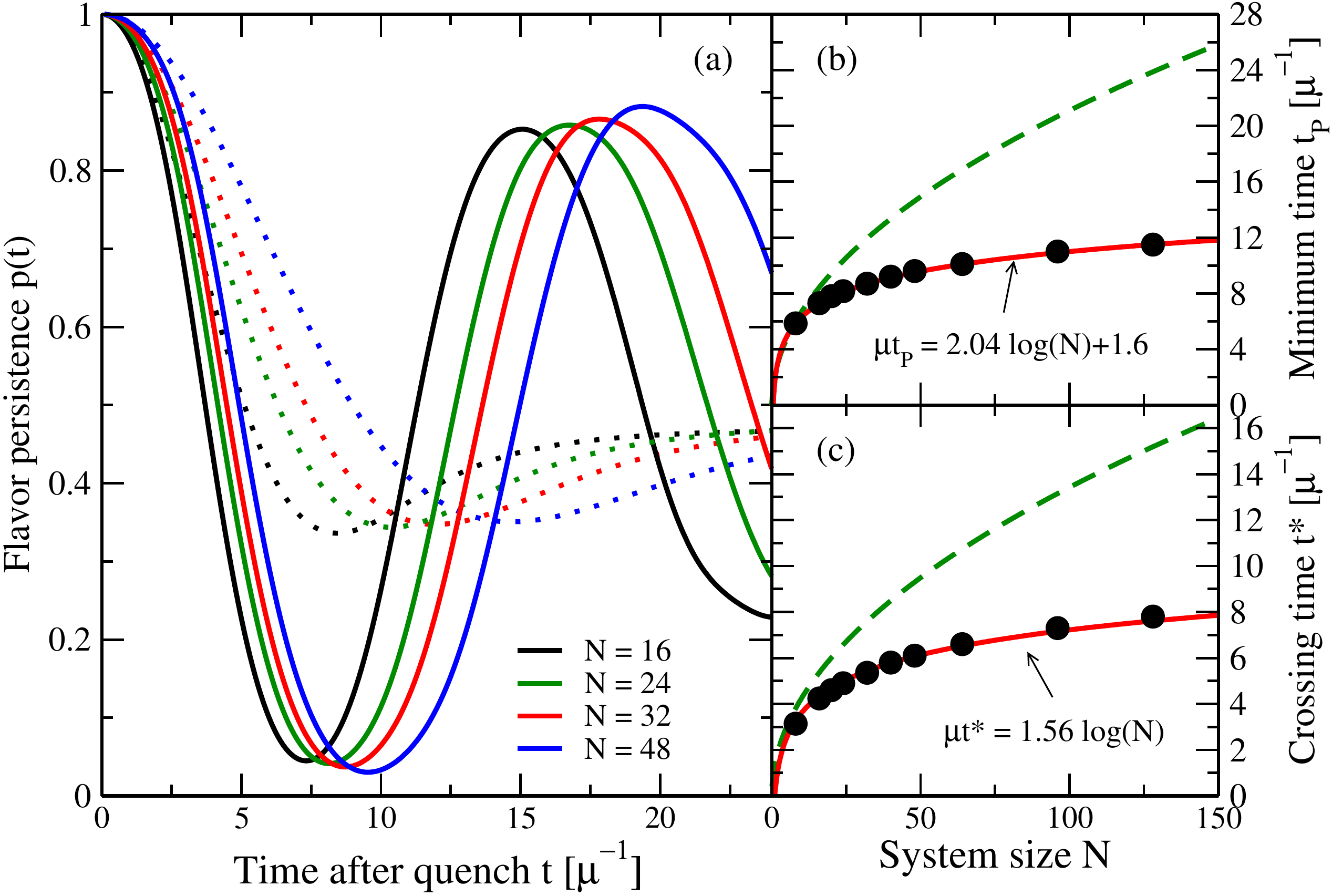}
 \caption{(Color online) The main panel (a) shows the flavor persistence $p(t)$ for different sizes of the neutrino system with $N$ given by: $16$ (black line), $24$ (green line), $32$ (red line) and $48$ (blue line). Also shown as dotted lines, with the same color, the results for $p(t)$ obtained with the single angle approximation in Sec.~\ref{sec:sa}. The right panels show the system size dependence of the time $t_P$ to reach the minimum of the persistence (panel (b)) and the crossing time $t^*$ of the Loschmidt echoes (panel (c)). The green dashed lines are the best fit for the results with the single angle approximation.}
\label{fig:2b_pers}
\end{figure}

Here we study in some detail the quench used in the original paper by Bell et al. in Ref.~\cite{Bell2003} (denoted by the $BRS$ arrow in Fig.~\ref{fig:2b_quench}) and comment on the qualitative differences with the single angle case explored in the previous section. Further exploration of the interplay between the equilibrium phase boundaries displayed in Fig.~\ref{fig:2b_phases} and the presence of a DPT would be very interesting. However, since in order to describe neutrino interactions we are not allowed to break the $SU(2)$ invariance explicitly, we cover here only the simpler case needed to explain the findings of Refs.~\cite{Bell2003,sawyer2004classical} and proceed in the next section to consider instead the $SU(2)$-invariant problem considered already in Ref.~\cite{Roggero2021A} which shows similar features.

We start by looking at both the time evolution of the flavor persistence $p(t)$ and the crossing time of the two Loschmidt echoes from Eq.~\eqref{eq:two_echoes}. The main panel of Fig.~\ref{fig:2b_pers} shows the flavor persistence $p(t)$ for various system sizes (solid lines) together with the equivalent result in the single angle approximation from the previous section (dotted lines). It is clear that flavor evolution happens much faster in the BRS quench, with sustained oscillations for long times. The frequency of these oscillations, as measured by the time $t_P$ to reach the first minimum, follows the fast time scale $\tau_F=\mu^{-1}\log(N)$ as $\mu t_P=2.04(5)\log(N) + 1.6(1)$ (see panel (b) of Fig.~\ref{fig:2b_pers}). This is in agreement with the expectations from results presented in Refs.~\cite{Bell2003,sawyer2004classical} and much faster than in the single angle approximation (shown as the green dashed line in Fig.~\ref{fig:2b_pers}(b)) we studied above and in Ref.~\cite{Friedland2003}.

The Loschmidt echoes $\mathcal{L}_0(t)$ and $\mathcal{L}_1(t)$ are also found to cross at shorter time scales than those found in Sec.~\ref{sec:sa}. The results for the crossing time $t^*$ as a function of the system size $N$ are presented in panel (c) of Fig.~\ref{fig:2b_pers} and again follow the fast time scale with $t^*/\tau_F = 1.56(4)$.

\begin{figure}
 \centering
 \includegraphics[width=0.49\textwidth]{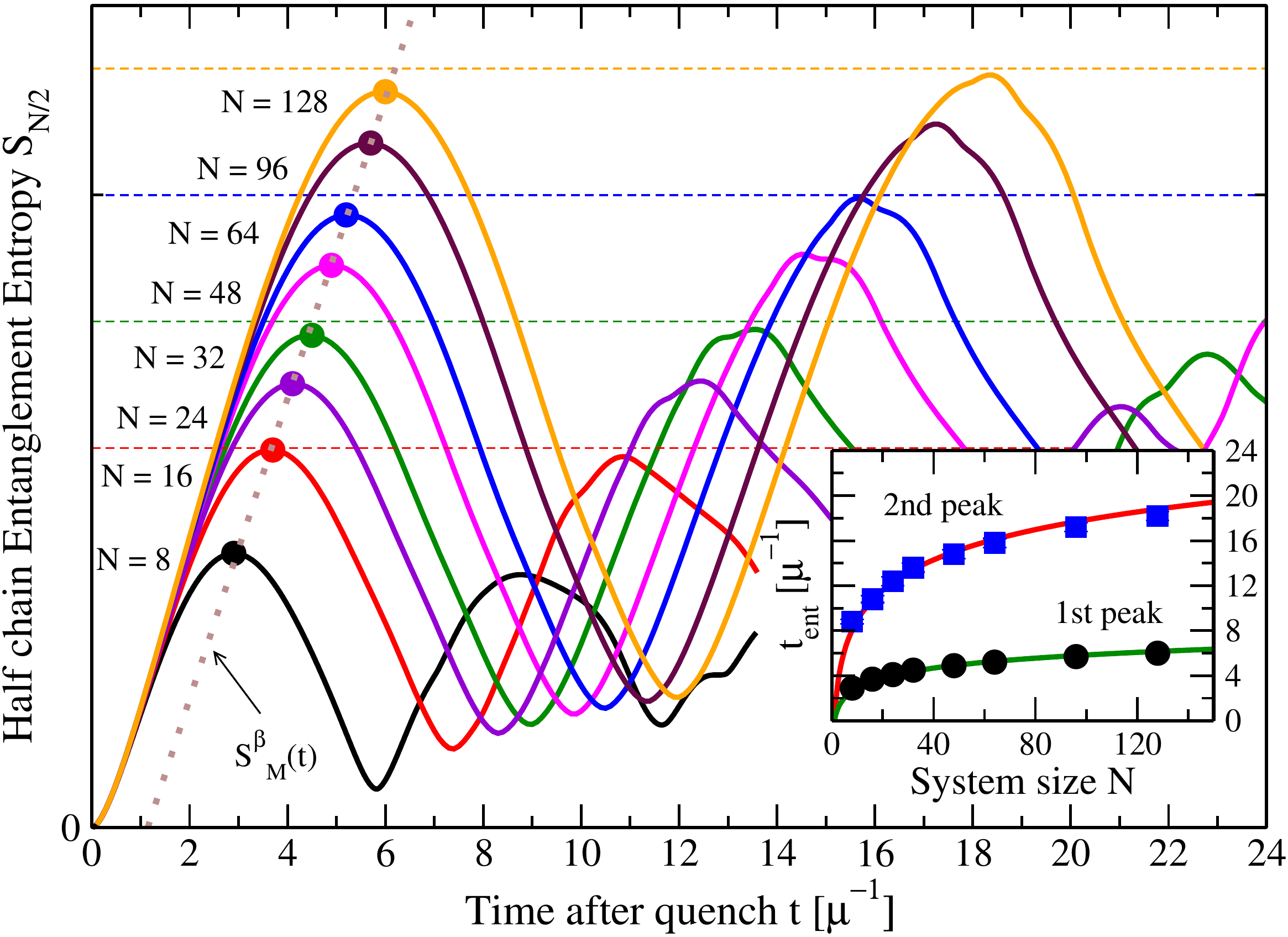}
 \caption{(Color online) Half chain entanglement entropy $S_{N/2}$ for different system sizes $N=8-128$ (solid lines). The horizontal dashed lines correspond to $S_{N/2}=S_{max}$ as in Fig.~\ref{fig:sa_entropy}. The inset shows the evolution in system size of both peaks, together with logarithmic fits. The brown dotted line in the main panel is the "case $\beta$" fit from Eq.~\eqref{eq:smaxoft}.  }
\label{fig:2b_entropy}
\end{figure}

The stark difference with the single angle case can also be observed in the evolution of the half-chain entanglement entropy $S_{N/2}$ defined in Eq.~\eqref{eq:entropy}. The main panel of Fig.~\ref{fig:2b_entropy} shows the entanglement entropy for different system sizes $N=8,16,24,32,48,64,96,128$ (solid lines in the main panel). The behavior in this case is qualitatively different from the results shown in Fig.~\ref{fig:sa_entropy} for the single angle approximation: the entanglement entropy itself oscillates in time, reaching values as high as $S_{max}$ (dashed lines in Fig.~\ref{fig:2b_entropy}) multiple times. In the results shown in Fig.~\ref{fig:2b_entropy} we see two distinct peaks whose times scale with the fast time scale $\tau_F$ as $t^1_{ent}/\tau_F=1.3(1)$ and $t^2_{ent}/\tau_F=3.9(1)$ respectively (these fits are shown in the inset of Fig.~\ref{fig:2b_entropy} as continuous lines). To corroborate these findings we also shown in the main panel is the "case $\beta$" fit from Eq.~\eqref{eq:smaxoft} which very accurately matches the evolution of the entropy maximum.

The results shown in this section were obtained using $\Gamma=0$ as in the original model from Ref.~\cite{Bell2003} but we confirmed the presence of the same logarithmic time scale also for larger values up to $\Gamma\approx0.7$ as observed also in previous work as reported in Ref.~\cite{sawyer2004classical}. The original model from Ref.~\cite{Bell2003} also used a more complex angular distribution than the two beam geometry employed here and in Ref.~\cite{sawyer2004classical}, unfortunately, due to the explicit $N$ dependence of the angular distribution used there, it was not possible to obtain a smooth extrapolation in system size as we have done with the other models in this work. We have found in a few selected cases at fixed $N$ that, with our initial state $\ket{\Psi_0}$, more complex angular dependence actually slows down the dynamics as compared to the two beam geometry. This effect is likely due to frustration of some of the interaction terms and in future work we plan to assess more quantitatively the role of multi-angle effects by using model geometries that have a well-defined scaling with system size.

\section{Intermediate density regime}
\label{sec:intdens}

The fast flavor oscillations observed in the models of the previous section are unfortunately not directly relevant to neutrino physics since the correct Hamiltonian is $SU(2)$ flavor invariant also in the general case. The previous result, however, points to the fact that oscillations at the time scale $\tau_F$ can appear when one crosses a quantum critical point and we have a DPT in the quantum quench. By tuning appropriately the one body part of the forward-scattering Hamiltonian in Eq.~\eqref{eq:fs_hamilt} we can orchestrate this to happen also in a physically relevant scenario closely related to the model used in describing bipolar collective oscillations (see eg.~\cite{Hannestad2006,Duan2006b}).

In this section we will consider the same model we introduced in the companion paper~\cite{Roggero2021A} where the system is still decomposed in the two beams $A$ and $B$ but now with two different energies
\begin{equation}
\begin{split}
H_{ID} &= -\frac{\omega_A}{2}\sum_{i\in\mathcal{A}}\sigma_i^z-\frac{\omega_B}{2}\sum_{i\in\mathcal{B}}\sigma_i^z + \frac{\mu}{2N}\sum_{i<j}\vec{\sigma}_i\cdot\vec{\sigma}_j\;,
\end{split}
\end{equation}
where we have also used the single angle approximation for the coupling matrix $\mathcal{J}_{ij}$ in the interaction. The Hamiltonian commutes with the z component of the total flavor spin $J_z=Z_A+Z_B$ and, given our initial state $\ket{\Psi_0}$, it's expectation vale remains zero at all times. Using spin operators for the neutrinos in the two beams and denoting the spin difference along the z axis as $D_{z}=Z_B-Z_A$, we can write the full Hamiltonian as (cf.~\cite{Roggero2021A})
\begin{equation}
\label{eq:hidr}
H_{ID} = \frac{\mu}{N} J^2 + \delta_\omega D_z \;
\end{equation}
where we introduced $\delta_\omega=(\omega_A-\omega_B)/2$ for the energy difference between the two beams and dropped an irrelevant constant. The equilibrium phase diagram depends on the sign of the two body interaction $\mu$:
\begin{itemize}
    \item for a ferromagnetic coupling $\mu<0$, there is a second order transition at $\delta_\omega=\pm|\mu|$ between polarized phases with $\langle D_z\rangle=\mp N/2$ and a broken phase with ferromagnetic order in the xy plane~\cite{Vidal2004c}.
    \item for an anti-ferromagnetic coupling $\mu>0$, the transition between gapped polarized phases is of first order and at $\delta_\omega=0$ instead~\cite{Vidal2004}.
\end{itemize}

On the other hand, the Loschmidt echo Eq.~\eqref{eq:loch} characterizing a DPT is invariant upon inversion of the full Hamiltonian $H_{ID}\rightarrow-H_{ID}$ and we can therefore expect the dynamical phase diagram to display features of both cases above and depend instead only on the relative sign of the two couplings constants $\mu$ and $\delta_\omega$.

This is indeed the case as shown in the results presented in Ref.~\cite{Roggero2021A} which we briefly summarize here. Using energy conservation together with the known initial state $\ket{\Psi_0}$ whose energy expactatin value reads
\begin{equation}
E_0 = \langle\Psi_0\lvert H_{ID}\rvert \Psi_0\rangle = \frac{\mu}{2}+\delta_\omega \frac{N}{2}\;,
\end{equation}
we can express the instantaneous value of the total angular momentum as a function of the staggered spin polarization $D_z$ as follows
\begin{equation}
\langle J^2(t)\rangle = \frac{N}{2}\left(1+2\frac{\delta_\omega}{\mu} \left(\frac{N}{2}-\langle D_{z}(t)\rangle\right)\right)\;,
\end{equation}
with initial conditions $\langle J^2(0)\rangle=\langle D_{z}(0)\rangle=N/2$. As was show in the accompanying paper~\cite{Roggero2021A}, this relation between the total angular momentum and the flavor asymmetry in the two beams is sufficient to characterize qualitatively the entire out-of-equilibrium dynamics. For completeness we provide a more complete derivation of those results with more details in the following.

In the case where the energy asymmetry $\delta_\omega/\mu<0$ is negative, the total spin, which starts already at a relatively small value, can only decrease further during time evolution. Since the operator $J^2$ is positive semi-definite this introduces a constraint on the fluctuations that $D_{z}$ can experience, in particular
\begin{equation}
\langle D_{z}(t)\rangle\bigg|_{\delta_\omega/\mu<0} \geq \frac{N}{2} -\left|\frac{\mu}{2\delta_\omega}\right|\;,
\end{equation}
and the change in polarization per spin vanishes in the thermodynamic limit. This suggests that for $\delta_\omega/\mu<0$ the system experiences negligible flavor evolution and is always close to the initial state, this was called the frozen phase in Ref.~\cite{Roggero2021A}. In the opposite limit $\delta_\omega/\mu > 0$ instead, the fluctuations become parametrically small at low densities (corresponding to $\delta_\omega/\mu \gg 1$ ) but remain finite also in the $N\gg1$ limit
\begin{equation}
\langle D_{z}(t)\rangle\bigg|_{\delta_\omega/\mu>0} \geq \frac{N}{2}\left(1 -\frac{\mu}{2\delta_\omega}\right)\;.
\end{equation}
This inequality provides a nontrivial bound on the spin, or flavor, fluctuations only for large $\delta_\omega > \mu/4$. Based on the discussion of the equilibrium phase diagram of this model, we expect to find the system in the gapped polarized phase, with $\langle D_z\rangle$ large and positive, for sufficiently large $\delta_\omega/\mu$ values. An estimate for the transition can be obtained by considering the minimum value of $\delta_\omega/\mu$ for which the first order fluctuations preserve the sign of the order parameter. This can be obtained by ensuring
\begin{equation}
\langle D_{z}(t)\rangle - \sqrt{\text{Var}[D_z](t)} > 0\;,
\end{equation}
with $\text{Var}[D_z](t)=\langle D^2_{z}(t)\rangle-\langle D_{z}(t)\rangle^2$ the variance of $D_z$. Using the fact that $\langle J_z\rangle = \langle J^2_z\rangle=0$ for our initial state, we can find the following upperbound on the variance
\begin{equation}
\begin{split}
\label{eq:var_bound}
\text{Var}[D_z](t) &= 2\left(\langle Z_A^2\rangle+\langle Z_B^2\rangle-\langle Z_A\rangle^2-\langle Z_B\rangle^2\right)\\
&= 2\left(\langle Z_A^2\rangle+\langle Z_B^2\rangle\right)-\langle D_z(t)\rangle^2\\
&\leq \frac{N^2}{4}-\langle D_z(t)\rangle^2\;.
\end{split}
\end{equation}
We therefore expect the system to be in the polarized phase and experience little flavor evolution when
\begin{equation}
\langle D_z(t)\rangle > \frac{N}{2\sqrt{2}}\quad\Rightarrow\quad \frac{\delta_\omega}{\mu} > \frac{1}{2-\sqrt{2}}\approx 1.7\;,
\end{equation}
and possibly at somewhat smaller values due to the bound Eq.~\eqref{eq:var_bound} being not tight.

Finally, in the regime $0<\delta_\omega/\mu\leq 1/4$ the total spin $J^2$, and correspondingly the flavor difference $D_{z}$, can experience strong fluctuations bounded by
\begin{equation}
\frac{N}{2}\leq \langle J^2(t)\rangle\bigg|_{0\leq\delta_\omega/\mu\leq1/4} \leq \frac{N}{2}\left(1+ 2\frac{\delta_\omega}{\mu} N \right)\;.
\end{equation}

As expected from this qualitative discussion, the dynamical phase diagram delineated above corresponds to a combination of the equilibrium phase diagrams of both the ferromagnetic and anti-ferromagnetic cases, with the exception that the transition at large $\delta_\omega/\mu$ appears shifted to larger values than $\delta_\omega/\mu=1$. 

\begin{figure}[b]
 \centering
 \includegraphics[width=0.45\textwidth]{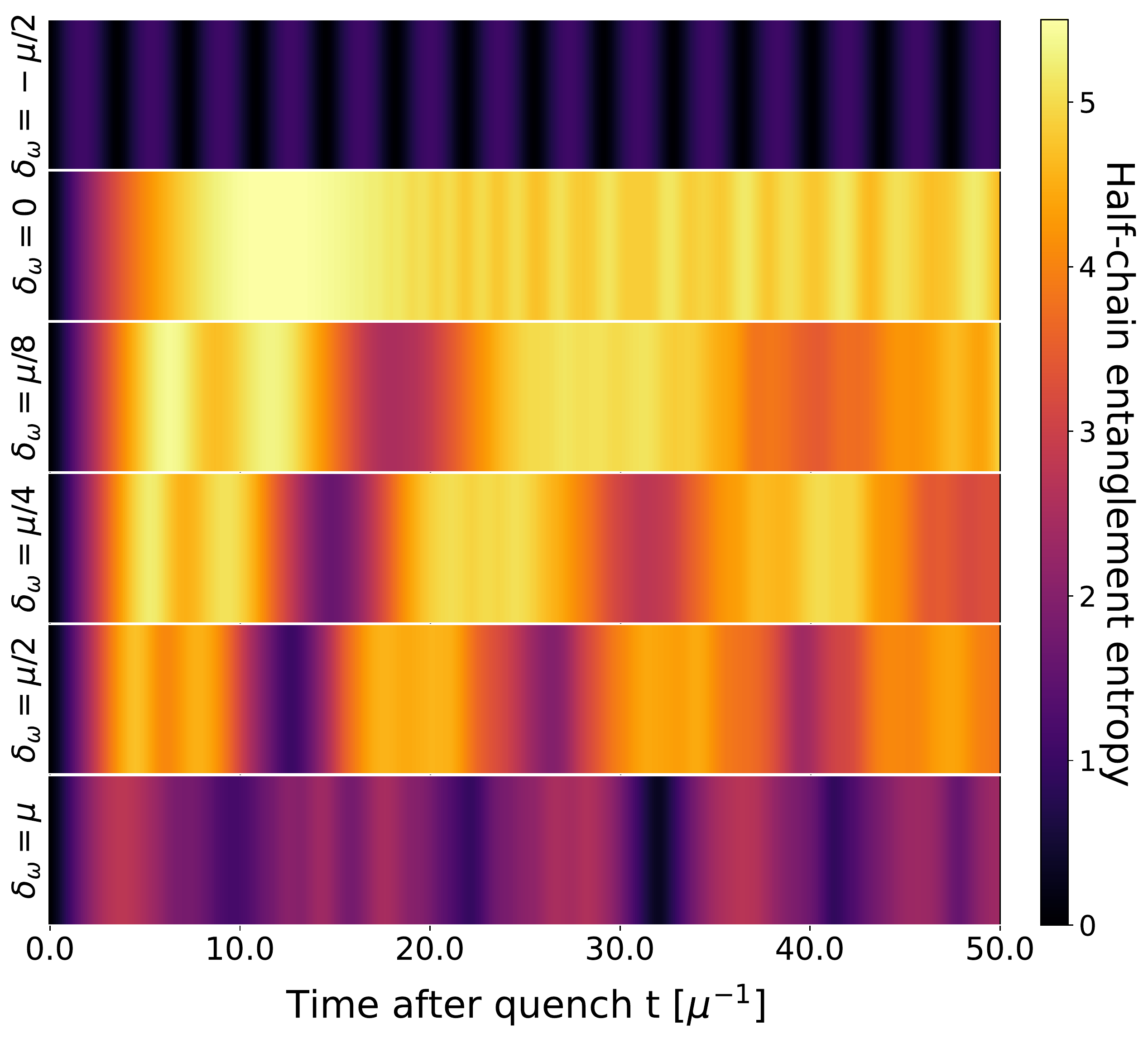}
 \caption{(Color online) Half-chain entanglement entropy for a system with $N=96$ neutrino amplitudes as a function of time for six values of the energy asymmetry parameter $\delta_\omega/\mu$ (from top to bottom): $-0.5,0.0,0.125,0.25,0.5,1.0$. }
\label{fig:sev}
\end{figure}

As shown also in Ref.~\cite{Roggero2021A}, the presence of these different dynamical phases is directly visible in the time evolution of the half-chain entanglement entropy for different values of the one body energy asymmetry $\delta_\omega/\mu$. In the frozen phases for either $\delta_\omega/\mu<0$ or $\delta_\omega/\mu\gtrapprox1$ the entanglement entropy remains small with a maximum value independent of system size. For negative energy asymmetry $\delta_\omega/\mu$ the entropy experiences fast oscillations which bring $S_{N/2}$ back to zero periodically. This is shown in the top panel of Fig.~\ref{fig:sev} showing the evolution of the half-chain entropy for a system of $N=96$ neutrino amplitudes across the different dynamical phases. In the gapless region $0<\delta_\omega/\mu\lesssim1$ the entanglement entropy shows strong fluctuations as a function of time, with maximum values close to $S_{max}=\log_2(N/2)$ and monotonically decreasing for increasing one-body energy asymmetry (see also Fig.3 of Ref.~\cite{Roggero2021A}). The special case $\delta_\omega/\mu=0$ matches the behavior presented in Fig.~\ref{fig:sa_entropy} above, with a peak at $t_{ent}\propto\mu^{-1}\sqrt{N}$ and small fluctuations at late times.
The scaling of time scales in the half-chain entropy for the unstable region $0<\delta_\omega/\mu\lesssim1$ shows a logarithmic behavior as expected from the presence of a DPT into a gapless phase, similarly to what we have found for the $SU(2)$-broken model in Sec.\ref{sec:su2b}.

In order to establish a closer connection to dynamical phase transitions as defined in the previous sections, we now consider the evolution of the Loschmidt echo $\mathcal{L}(t)$ for different values of the asymmetry parameter $\delta_\omega$ in all three dynamical phases. 
Contrary to the situation in Sec.~\ref{sec:sa} and Sec.~\ref{sec:su2b}, the initial Hamiltonian we consider here (namely $H_{ID}$ with $\delta_\omega<0$ and $\mu=0$) has a unique ground-state. For all quenches with $\delta_\omega\neq 0$ considered in this section, we have always found $\mathcal{L}_1(t)\approx0$ for large system sizes and a DPT will not appear as a crossing of echoes as before, but instead as sharp peaks in the Loschmidt rate $\lambda(t)$ defined in Eq.~\eqref{eq:loch_rate} above.

\begin{figure}[t]
 \centering
 \includegraphics[width=0.49\textwidth]{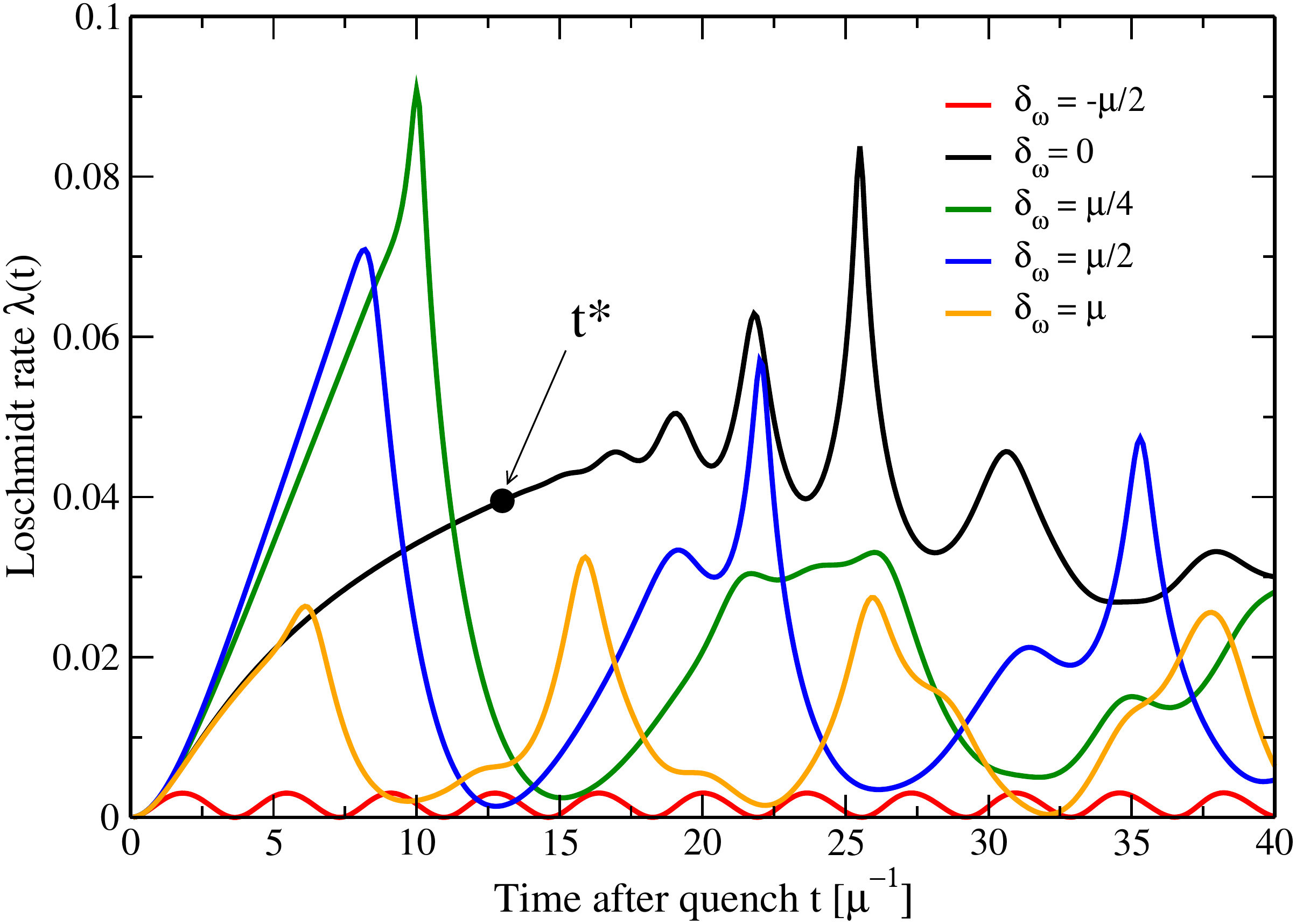}
 \caption{(Color online) Loschmidt rate in a system with $N=96$ neutrino amplitudes initialized in $\ket{\Psi_0}$ and quenched at different values of the one body asymmetry parameter: the red line corresponds to a negative value $\delta_\omega=-\mu/2$, the black line contains only the two body potential as in Fig.~\ref{fig:sa_echoes} and the green, blue and orange lines correspond to positive asymmetries $\delta_\omega=(0.25,0.5,1.0) \mu$ respectively.}
\label{fig:losc}
\end{figure}

This is illustrated in of Fig.~\ref{fig:losc} where the Loschmidt rate $\lambda(t)$ is shown for different values of $\delta_\omega$ in a system of $N=96$ neutrino amplitudes. The purely two-body case at $\delta_\omega=0$ has a DPT generated by crossing Loschmidt echoes at $t=t^*$ (shown as a dot in Fig.~\ref{fig:losc}), followed by additional sharp features at later times. For negative values of $\delta_\omega$, in the frozen phase, the rate $\lambda(t)$ remains smooth at all times, while for positive $\delta_\omega$ sharp features start to appear at even shorter times than the $t^*$ crossing time and a DPT can occur in the system. Obtaining an estimate for the critical time where a DPT might occur in this case is complicated by it's expected evolution with system size, in parallel to the case $\delta_\omega=0$ considered in Sec.~\ref{sec:sa} above. This has prevented a reliable extraction of a unique critical time $t^*$ in the unstable region $0<\delta_\omega/\mu\lesssim1$ using results up to $N=128$ and a single value for the time-step of the evolution (here we used $0.05\mu^{-1}$ as in Ref.~\cite{Roggero2021A}).
This observation highlights the usefulness of entanglement measures such as the half-chain entropy as a more robust indicator of the presence of qualitative changes in the dynamical phase of a many-body quantum system. Future explorations employing either semi-classical approaches, like those used for instance in \cite{Zunkovic2016}, or specialized simulations exploiting more directly symmetries of the system, are expected to be able to clarify the role of fidelity measures as the Loschmidt echo in characterizing the different dynamical phases found in models of neutrino flavor evolution.

\section{Summary and Conclusions}
\label{sec:conc}

The presence of collective oscillations in the dynamical evolution leading to neutrino flavor transport has long been recognized as an important effect in describing the dynamics of astrophysical environments like supernovae and the early universe~\cite{PANTALEONE1992,Pantaleone92,Pastor2002,Pastor2002B}. Early explorations by Sawyer and coworkers~\cite{Bell2003,sawyer2004classical,Sawyer2005} suggested that quantum correlations, in the many-body spin system corresponding to a neutrino cloud, could lead to a coherent speed-up of collective oscillations, with possibly important consequences for the dynamics of these environments. This idea, which invites caution on the interpretation of results for the neutrino flavor evolution obtained using mean-field approximations (which neglect quantum entanglement), has been challenged in the past by presenting counter-examples in solvable models where the qualitative prediction of the mean-field are matched by the exact solution~\cite{Friedland2003,Friedland2006}. The absence of entanglement in the neutrino dynamics more generally has also being argued as a justification for the mean-field approach to the problem~\cite{Friedland2003b}. This debate has recently re-emerged thanks to works like Ref.~\cite{Cervia2019} and Ref.~\cite{Rrapaj2020} which showed that entanglement is indeed produced when solving exactly the many-body neutrino problem encoded in the forward scattering Hamiltonian of Eq.~\eqref{eq:fs_hamilt} and it's time-dependent generalizations. The explored systems were however too small ($N=\mathcal{O}(10)$) to draw general conclusions applicable to the large collections of neutrino amplitudes needed for realistic simulations.

Exploiting the expectation that the entanglement entropy is unlikely to grow too large in these many-body systems, due to the infinite range of interactions in the spin model of Eq.~\eqref{eq:fs_hamilt}, the present work extends the idea presented in the companion paper Ref.~\cite{Roggero2021A} to use a Matrix Product State (MPS) representation in order to efficiently describe the neutrino wave-function as it evolves from an initial product state. As explained in more detail in Ref.~\cite{Roggero2021A}, this approach is ideal for low levels of bipartite entanglement in the system and allows to easily simulate systems with $\approx100$ neutrino amplitudes with modest computational resources. This simulation strategy is used here with two main goals, the first one was to validate the early small scale simulations by Sawyer et al.~\cite{Bell2003,sawyer2004classical} which, correctly, predicted flavor evolution to occur (in their model) at a fast time-scale $\tau_F\approx\mu^{-1}\log(N)$. This shows that indeed many-particle neutrino interactions cause a novel coherent effect not captured by the mean-field approximation. A similar effect is also found in the more familiar bipolar oscillations described in detail in Ref.~\cite{Roggero2021A} and Sec.~\ref{sec:intdens} of the present work. The second goal was to explain the presence of this fast time scale as being generated by an underlying Dynamical Phase Transition. This observation explains the absence of the effect in the exactly solvable models discussed in Refs.~\cite{Friedland2003,Friedland2006} and provides a more direct link between the presence of coherently-enhanced flavor oscillations and non-negligible levels of entanglement in the many-body state generated by the dynamics.

The work presented here and in the accompanying paper Ref.~\cite{Roggero2021A} opens the way to accurate many-body simulation of the full quantum dynamics of neutrino flavor transport with controllable errors. The use of entanglement-efficient methods, like the MPS representation used here, will allow for the first time a more direct comparison with popular approximation methods working in the mean field for large system sizes. This will be critical to allow for the inclusion of rich energy/angle distributions and avoid the limitations of special symmetric points like the model studied in Ref.~\cite{Friedland2003} and covered in Sec.~\ref{sec:sa} of the present work. Possible failures of this program would be associated to situations where the entanglement entropy grows substantially with system size. The identification of the parameter regimes where this happens would shed light on potentially interesting candidates to study using quantum computing devices as recently explored in Ref.~\cite{Hall2021}. Finally, a better understanding of the dynamical phase diagram of neutrino models, as the one described in Eq.~\eqref{eq:fs_hamilt} and it's generalization to the full 3 flavor case, would help identify the conditions (beyond linear stability analysis) required for collective oscillations to appear in complex environments like supernovae explosions by an appropriate analysis of simulation results. Work is ongoing to extend the results presented in this work to more realistic conditions in order to better asses the impact of entanglement in astrophysical settings with large neutrino densities.

\begin{acknowledgments}
I want to thank Joseph Carlson, Vincenzo Cirigliano, Huaiyu Duan, Joshua Martin, Ermal Rrapaj and Martin Savage for the many useful discussions about the subject of this work. This work was supported by the Institute for Nuclear Theory under U.S. DOE grant No. DE-FG02-00ER41132, by the InQubator for Quantum Simulation under U.S. DOE grant No. DE-SC0020970 and by the Quantum Science Center (QSC), a National Quantum Information Science Research Center of the U.S. Department of Energy (DOE).
\end{acknowledgments}
 
%

\appendix

\section{Details on equilibrium phase diagrams}
\subsection{Phase diagram for single angle model}
\label{app:phase_sa}
In the quantum quench relevant for the single angle case studied in Sec.~\ref{sec:sa} the full Hamiltonian reads
\begin{equation}
H = \frac{\mu}{N}J^2  + \frac{\nu}{N} Z_AZ_B\;,
\end{equation}
with $\mu$, $\nu$ positive constants.
As discussed in the main text, in the limit $\nu=0$ the system is gapless and the groundstate has zero total angular momentum and zero energy. In the limit $\mu=0$ instead, the system has two degenerate ground-states which, in the angular momentum basis $\ket{s_A,m_A}\otimes\ket{s_B,m_B}$ of the two set of spins with total angular momenta $\vec{S}_A$ and $\vec{S}_B$, we can write as
\begin{equation}
\begin{split}
\ket{GS_0} &= \ket{\frac{N}{4},-\frac{N}{4}}\otimes\ket{\frac{N}{4},\frac{N}{4}}\\
\ket{GS_1} &= \ket{\frac{N}{4},\frac{N}{4}}\otimes\ket{\frac{N}{4},-\frac{N}{4}}\;.
\end{split}
\end{equation}
In these configurations the system has an anti-ferromagnetic order across beams characterized by $\langle Z_AZ_B \rangle=-N^2/16$. In the gapless phase the order parameter is zero. The expectation value of the full Hamiltonian in either of the anti-ferromagnetic states reads
\begin{equation}
\langle GS_k \lvert H\rvert GS_k\rangle = \frac{\mu}{2} - \nu\frac{N}{16}\;,
\end{equation}
and becomes negative for a sufficiently large antiferromagnetic coupling $\nu>8\mu/N$. In the thermodynamic limit we expect the critical point to be at $\nu=0$ for any $\mu>0$. As we will see in a more general case below, if we allow $\nu$ to become negative other phases emerge.

\subsection{Phase diagram of the two-beam model}
\label{app:tb_phase_diag}

In this section we provide more details on the calculation of the mean-field phase diagram presented in Fig.~\ref{fig:2b_quench} of the main text. This corresponds to the ground-state phase diagram of the following Hamiltonian (cf. Eq.\eqref{eq:ham_tb})
\begin{equation}
\begin{split}
\label{eq:two_beams_diag}
H &= \frac{\Gamma}{N}\left[J_A^2 + J_B^2 + \left(\Delta-1\right) \left({Z_A}^2+{Z_B}^2\right)\right]\\
&+\frac{2}{N}\left[\vec{J}_A\cdot\vec{J}_B +\left(\Delta-1\right)Z_AZ_B\right]\;,
\end{split}
\end{equation}
with a positive coupling constant $\Gamma=\mathcal{J}_{AA}/\mathcal{J}_{AB}$.

The order parameters of interest here are the average staggered magnetizations of the two beams
\begin{equation}
\begin{split}
M^{XY}_{AB} &=\frac{1}{N} \langle X_AX_B\rangle+\langle Y_AY_B\rangle\\
M^{Z}_{AB} &= \frac{1}{N} \langle Z_AZ_B\rangle\\
M^{V}_{AB} &= \frac{1}{N} \langle \vec{J}_A\cdot\vec{J}_B\rangle = M^{XY}_{AB}+M^{Z}_{AB}
\end{split}
\end{equation}

We start the discussion of the equilibrium phase diagram by considering first some special cases:
\begin{itemize}
    \item at the $SU(2)$ symmetric point, corresponding to $\Delta=1$, we have the following Hamiltonian
    \begin{equation}
    H_{\Delta=1} = \frac{1}{N}J^2+\frac{\Gamma-1}{N}\left(J_A^2 + J_B^2\right)\;.
    \end{equation}
    For $\Gamma<1$ the system is in an anti-ferromagnetic gapless phase characterized by $M^V_{AB} = -\frac{N}{16}$ and undefined values for $M^{XY}_{AB}$ and $M^Z_{AB}$. For $\Gamma>1$ we have instead a disordered gapless phase characterized by a vanishing order parameters $M^V_{AB}=M^{XY}_{AB}=M^Z_{AB} =0$. At the single angle point $\Gamma=1$, the three order parameters are undefined. Note that, when the initial state is $\ket{\Psi_0}$ from Eq.~\eqref{eq:init_state}, the resulting evolution is the same for any value of $\Gamma$ since $S_A^2$ and $S_B^2$ are conserved quantities.
    \item at the single angle point $\Gamma=1$ we have instead
    \begin{equation}
    \begin{split}
    H_{\Gamma=1} &= \frac{1}{N}J^2 + \frac{1}{N}(\Delta-1){Z_{tot}}^2\\
    &=\frac{1}{N}\left(X^2_{tot}+Y^2_{tot}\right) + \frac{1}{N}\Delta{Z_{tot}}^2\;,
    \end{split}
    \end{equation}
    with $Z_{tot}=Z_A+Z_B$ the total spin in the z direction (and similarly for $X_{tot}$ and $Y_{tot}$). The groundstate of this model for $\Delta<0$ is a (gapped) ferromagnet with $M^Z_{AB}=\frac{N}{16}$, for $\Delta\geq0$ the groundstates are the singlet states with zero total spin and with undefined order parameters. Given our initial state $\ket{\Psi_0}$, and the fact that $\left[Z_{tot},H_{\Gamma=1}\right]=0$, the time evolution is exactly equivalent to the the single angle case studied above for any value of $\Delta$.
    \item for collimated beams with $\Gamma=0$ we have simply
    \begin{equation}
    H=\frac{2}{N}\left[\vec{J}_A\cdot\vec{J}_B +\left(\Delta-1\right)Z_AZ_B\right]\;.
    \end{equation}
    For $\Delta>1$ the ground states are $\ket{\Psi_0}$ and the spin reversed partner $\ket{\Psi_1}$ introduced in Sec.~\ref{sec:sa} of the main text. The system has anti-ferromagnetic order with $M^Z_{AB}=-\frac{N}{16}$ and there is a finite energy gap to excited states. For $-1<\Delta<1$ the system is gapless with $M^Z_{AB}=0$, in fact we have a continuum of zero-energy modes polarized in the XY plane with $M^{XY}_{AB} = - \frac{N}{16}$. Finally, for $\Delta<-1$ the system is a ferromagnet along the Z direction with $M^Z_{AB}=\frac{N}{16}$ and $M^{XY}_{AB}=0$.
\end{itemize}

In order to get the rest of the phase diagram we will compare energies of the different phases in the mean field limit. Let's first rewrite the Hamiltonian as
\begin{equation}
\begin{split}
H =& \frac{\Gamma}{N}\left[J_A^2+J_B^2\right]+ \frac{\Gamma}{N}(\Delta-1)\left[{Z_A}^2+{Z_B}^2\right]\\
&+ \frac{2}{N}\left[X_AX_B+Y_AY_B\right]+\frac{2\Delta}{N}Z_AZ_B\;.
\end{split}
\end{equation}
The mean field states we will consider here are:
\begin{equation}
\begin{split}
\ket{\Phi_{FM}} &= \bigotimes_{i=1}^N \ket{\uparrow}\\
\ket{\Phi_{AFM}} &= \left(\bigotimes_{i=1}^{N/2} \ket{\uparrow}\right)\otimes\left(\bigotimes_{i=1}^{N/2} \ket{\downarrow}\right)\\
\ket{\Phi_{XY}} &= \left(\bigotimes_{i=1}^{N/2} \ket{+}\right)\otimes\left(\bigotimes_{i=1}^{N/2} \ket{-}\right)\\
\end{split}
\end{equation}
together with the disordered state $\ket{\Phi_{DIS}}$ with zero total spin in beam A and B. In the expression above we use the notation $\ket{\pm}$ to indicate the eigenstates of the Pauli X operator with positive and negative eigenvalue respectively. The corresponding expectation values for the energy in the full Hamiltonian Eq.~\eqref{eq:two_beams_diag} are
\begin{equation}
\begin{split}
E_{FM} &= \frac{\Gamma}{2}+\frac{N}{8}\Delta\left(\Gamma+1\right)\;,\\    
E_{AFM} &= \frac{\Gamma}{2}+\frac{N}{8}\Delta\left(\Gamma-1\right)\;,\\    
E_{XY} &= \frac{\Gamma}{2}+\frac{N}{8}\left(\Gamma-1\right)\;,\\
E_{DIS} &= 0\;.
\end{split}
\end{equation}

\begin{figure}
 \centering
 \includegraphics[width=0.49\textwidth]{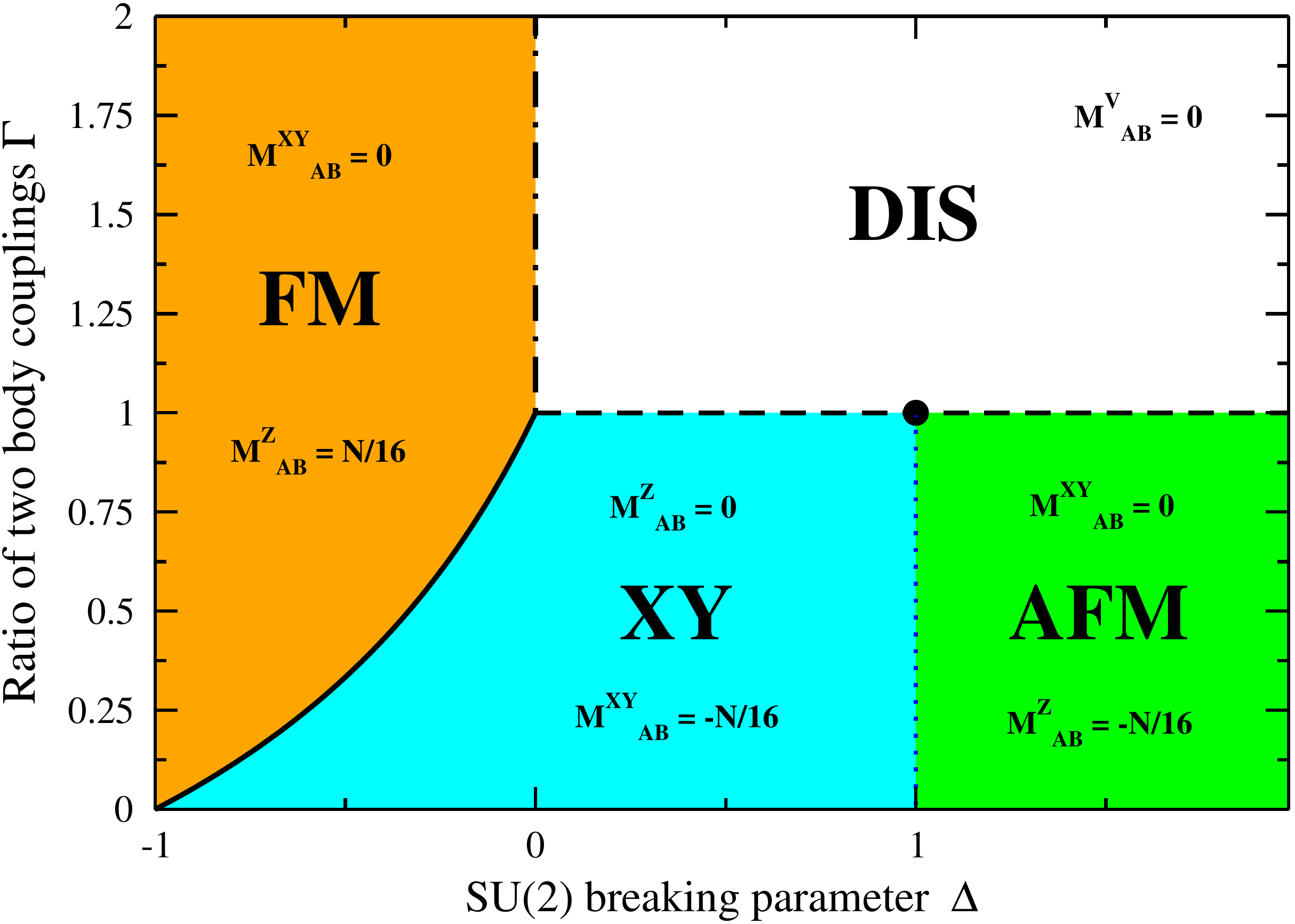}
 \caption{(Color online) Equilibrium phase diagram for the two beam model in Eq.~\eqref{eq:two_beams_diag}, see text for description of the phases and properties of the boundaries. The solid circle is the $SU(2)$ invariant model in the single angle approximation. Also shown are the values of the order parameters.}
\label{fig:2b_phases}
\end{figure}

The resulting phase diagram is depicted in Fig.~\ref{fig:2b_phases}. Along the critical lines separating the different phases we have the following
\begin{itemize}
    \item boundaries between $AFM$ and $DIS$ and between $XY$ and $DIS$ (dashed black curve in Fig.~\ref{fig:2b_phases}): all the order parameters are undefined due to the degeneracy of the spectrum for states with different values of the total spin in the two beams but zero total angular momentum.
    \item boundary between $AFM$ and $XY$ (dotted black curve in Fig.~\ref{fig:2b_phases}): the direction-independent magnetization takes the smallest value $M^{V}_{AB}=-\frac{N}{16}$ while the other two order parameters are undefined thanks to the $SU(2)$ invariance of the system. At the critical point for $\Gamma=1$ also $M^{V}_{AB}$ is undefined.
    \item boundary between $DIS$ and $FM$ (dash dotted black curve in Fig.~\ref{fig:2b_phases}): similarly to the boundary between $DIS$ and the other two ordered phases, all the order parameters can take values in $[-N/16,0]$.
    \item boundary between $FM$ and $XY$ (solid black curve in Fig.~\ref{fig:2b_phases}): the direction independent magnetization can take any value (both positive and negative) while $M^{XY}_{AB}\in[-N/16,0]$ and $M^{Z}\in[0,N/16]$.
\end{itemize}

\end{document}